\documentclass{article}

\usepackage{amssymb,amsmath}
\usepackage{fullpage}
\usepackage{url}
\usepackage{tikz}
\usetikzlibrary{petri} 
\usetikzlibrary{chains}

\begin{document}

\title{An example illustrating the imprecision \\
of the efficient approach for diagnosis of Petri nets 
\\via integer linear programming}
\author{Alban Grastien%
  \footnote{Alban Grastien is with the Optimisation Research Group of NICTA 
  and the Artificial Intelligence Group of ANU, Canberra, Australia.  
  E-mail: \texttt{alban.grastien@nicta.com.au}.}
}
\date{September 18, 2012}

\maketitle

\begin{abstract}
  This document demonstrates 
  that the efficient approach for diagnosis of Petri nets 
  via integer linear programming 
  may be unable to detect a fault 
  even if the system is diagnosable.  
\end{abstract}

\section{Introduction}  

The efficient approach for diagnosis of Petri nets 
via integer linear programming 
was proposed in 2009 by Basile et al. \cite{basile-etal::tac::09}.  
Their work represents a novel approach 
to diagnose discrete event systems \cite{cassandras-lafortune::99} 
modeled as a Petri net.  

The original work contains two typos \cite{basile-etal::misc::12}.  
In particular, the efficient approach comes at a cost, 
which is that a fault might come undetected 
even if the system is diagnosable.  
I illustrate this imprecision 
by presenting a diagnosable net system 
where the fault cannot be diagnosed using Basile et al.'s approach.  

It might be argued that the community was already aware 
that this approach is not precise.  
Dotoli et al. \cite{dotoli-etal::automatica::09} 
presented a similar yet precise approach, 
by removing the largest aspect of the efficiency in Basile et al. 
It seemed to me however that it was necessary to illustrate this point.  

\section{Efficient Diagnosis of Petri Net}

We assume that the reader is familiar with the Petri net formalism, 
and provide only a short definition.  
Formal definitions and further references 
can be found in Basile et al.'s paper \cite{basile-etal::tac::09}.  

A Petri net (cf.{} Fig.~\ref{fig::dablepetrinet}) 
consists in a set of places (represented by circles) 
and a set of transitions (represented by rectangles).  
A place contains a number of tokens (represented by dots) 
A number of edges connect places to transitions 
and transitions to places.  
A transition may trigger if all preceding places have a token; 
this consumes one token in each of these places 
and adds a token in each place that is a successor of the transition.  
A net system is a Petri net with a set of tokens.  

Some transitions are observable: their firing is known.  
Observable transitions are filled in black; 
unobservable transitions are empty.  
It is assumed that the sub-network 
without the observable transitions is cycle-free.  
One transition is faulty.  
It is unobservable.  

The purpose of diagnosis is the following.  
The system takes a sequence of transitions $s$ 
(under the liveness property, we assume 
that the system will never stop running, 
which means that $s$ can be assumed arbitrary long).  
The observation $o$ on the system is the projection of $s$ 
on the observable transitions 
(i.e., we remove from $s$ the unobservable transitions).  
Diagnosis consists in determining 
whether the faulty transition certainly/possibly occurred.  

I will not present diagnosis of Petri nets via integer linear programming 
as this is irrelevant for this discussion; 
I only need to say that this approach computes all the sequences $s'$ 
that can generate the observation $o$ 
and determine if all/some of these sequences 
contain the faulty transition.  
I will however discuss the efficient approach presented by Basile et al.  

There are two important aspects in this approach: 
\begin{enumerate}
\item 
Gven an observation $o = [t_1,\dots,t_n]$, 
the efficient approach computes the diagnosis 
of $o_1 = [t_1]$, of $o_2 = [t_1,t_2]$, \dots, 
and of $o_n = [t_1,\dots,t_n]$.  
If a fault is detected for some $o_i$, it is detected for $o$.  
\item 
A second important point about the efficient algorithm 
is that the order of the observed transitions is lost 
when the diagnosis of $o_k$ is performed.  
For instance, if $o_2 = [A,B]$, 
then the sequences $s'$ of transitions that are generated 
are those that generate $[A,B]$ or $[B,A]$.  
\end{enumerate}

\section{Example}

\begin{figure}[ht]
  \begin{center}
\begin{tikzpicture}[yscale=-1.1,thin,>=stealth,
  every transition/.style={fill,minimum width=1mm,minimum height=5mm},
  every place/.style={draw,thick,minimum size=6mm}]

  \node[place,tokens=1] (p1) at (4,1) {};
  \node[place,tokens=1] (p2) at (2,4) {};
  \node[place,tokens=1] (p3) at (6,4) {};
  \node[place] (p4) at (3,3) {};
  \node[place] (p5) at (5,3) {};
  \node[place] (p6) at (0,5) {};
  \node[place] (p7) at (5,5) {};
  \node[place] (p8) at (3,5) {};
  \node[place] (p9) at (8,5) {};
  \node[place] (p0) at (4,7) {};

  \node[transition,label=above:$f$,fill=none] at (0,2) {}
    edge [pre]  (p1)
    edge [post] (p6)
    edge [post] (p4)
  ;
  \node[transition,label=above:$u_1$,fill=none] at (3,2) {}
    edge [post] (p5)
    edge [post] (p6)
    edge [pre]  (p1)
  ;
  \node[transition,label=above:$u_2$,fill=none] at (8,2) {}
    edge [post] (p4)
    edge [post] (p9)
    edge [pre]  (p1)
  ;
  \node[transition,label=right:$A$] at (3,4) {}
    edge [pre]  (p2)
    edge [pre]  (p4)
    edge [post] (p8)
    edge [post] (p5)
  ;
  \node[transition,label=left:$B$] at (5,4) {}
    edge [pre]  (p3)
    edge [pre]  (p5)
    edge [post] (p7)
    edge [post] (p4)
  ;
  \node[transition,label=left:$D$] at (2,6) {}
    edge [pre]  (p6)
    edge [pre]  (p7)
    edge [pre]  (p8)
    edge [post] (p0)
  ;
  \node[transition,label=right:$C$] at (6,6) {}
    edge [pre]  (p7)
    edge [pre]  (p8)
    edge [pre]  (p9)
    edge [post] (p0)
  ;
  \node[transition,label=below:$E$] at (4,8) {}
    edge [pre,bend left]  (p0)
    edge [post,bend right] (p0)
  ;

\end{tikzpicture}
  \end{center}
  \caption{A diagnosable net system.}
  \label{fig::dablepetrinet}
\end{figure}
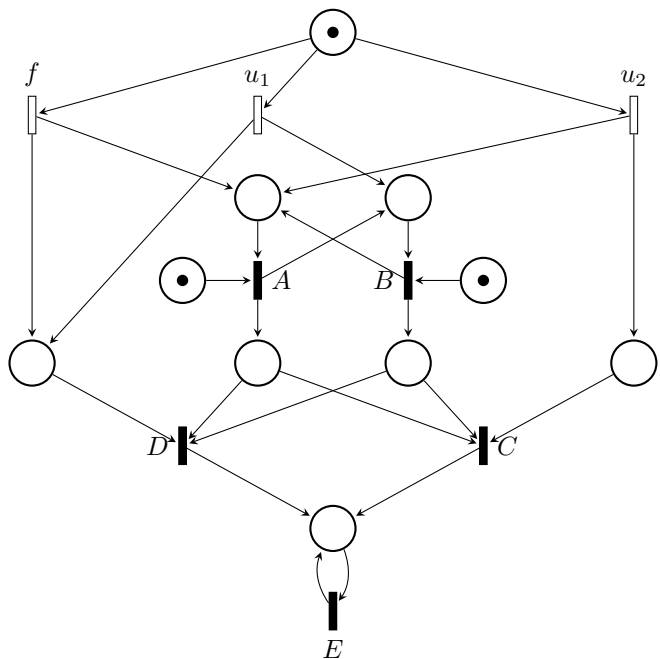

I now present the example that shows that the efficient approach 
is not precise.  
The net system is presented on Fig.~\ref{fig::dablepetrinet}.  
The observable transitions are $A$, $B$, $C$, $D$, and $E$; 
the unobservable transitions are $f$, $u_1$, and $u_2$, 
where $f$ is the faulty transition.  

There are three possible sequences of transitions: 
\begin{itemize}
\item $u = [f  ,A,B,D,E,E,\dots]$, 
\item $v = [u_1,B,A,D,E,E,\dots]$, and 
\item $w = [u_2,A,B,C,E,E,\dots]$.  
\end{itemize}
Clearly, the system is diagnosable\footnote{
Diagnosability means that a fault can always be detected.}, 
since a fault occurred iff $A$ is observed before $B$, 
and $D$ is observed.  

Assume now that $u$ takes place 
and $o = [A,B,D,E,E,\dots]$ is observed.  
I provide the list of explanations $s'$ 
associated with each observation $o_i$ 
(remember that is order between the observed transitions 
is dropped when the diagnosis of $o_i$ is performed): 
\begin{itemize}
\item $o_0 = []$: $s' = []$ 
\item $o_1 = [A]$: $s'_1 = [f,A]$, $s'_2 = [u_2,A]$ 
\item $o_2 = [A,B]$: $s'_1 = [f,A,B]$, $s'_2 = [u_2,A,B]$, 
  $s'_3 = [u_1,B,A]$
\item $o_3 = [A,B,D]$: $s'_1 = [f,A,B,D]$, $s'_2 = [u_1,B,A,D]$, 
\item $o_4 = [A,B,D,E]$: $s'_1 = [f,A,B,D,E]$, $s'_2 = [u_1,B,A,D,E]$, 
\item etc. 
\end{itemize}

Clearly, the fault is never diagnosed, 
because there is always another explanation 
that includes no fault.  

\paragraph*{Conclusion:}

This demonstrates that the efficiency of Basile et al.'s approach 
comes at the cost of precision.  
Despite this negative result, 
I see this approach as very interesting, 
in particular in connection with Petri net specific techniques 
(i.e., the reduction to integer linear programming).  

I think an interesting follow-up of this result 
would be to devise a precision testing 
that could decide whether a specific net system 
can be precisely diagnosed with the efficient approach.  
Even better would be to find sufficient restrictions 
on the Petri net (similar to the Petri net safety).  

\section*{Acknowledgments}
NICTA is funded by the Australian Government 
as represented by the Department 
of Broadband, Communications and the Digital Economy 
and the Australian Research Council 
through the ICT Centre of Excellence program.

\bibliographystyle{alpha}
\bibliography{bib}

\end{document}